\documentclass[a4paper]{spie}  

 \newcommand*\arcsec{\ensuremath{^{\prime\prime}}}

\usepackage{amsmath,amsfonts,amssymb}
\usepackage{graphicx}
\usepackage[colorlinks=true, allcolors=blue]{hyperref}
\usepackage{longtable}
\usepackage{eurosym}

\title{The MICADO first-light imager for the ELT \\The SCAO module AIT in France: from the $\beta$ flat configuration towards the flat and final configurations}

\author[a]{Yann Cl\'enet} 
\author[a]{Vincent Lapeyr\`ere}
\author[a]{Eric Gendron} 
\author[a]{Fabrice Vidal}
\author[a]{Fr\'ed\'eric Chapron} 
\author[b]{Mathieu Cohen}
\author[c]{Sylvain Guieu} 
\author[a]{Roderick Dembet} 
\author[e]{Eric Meyer} 
\author[a]{Arnaud Sevin}
\author[b]{Fanny Chemla}
\author[a]{Olivier Dupuis}
\author[a]{K\'evin Cloiseau} 
\author[b]{Gilles Fasola}
\author[a]{Karim Hussein}
\author[a]{Sylvain Isambert}
\author[b]{Julien Gaudemard}
\author[a]{Florian Ferreira}
\author[e]{Herv\'e Locatelli} 
\author[e]{Nicolas Gautherot} 
\author[e]{Antonin Poyard} 
\author[a]{Lahoucine Ghouchou}
\author[e]{Emmanuel Tisserand} 
\author[d]{Alexandre Blin}
\author[d]{Labasse Berthe}
\author[d]{Srilakshmi Ramalingam}
\author[a]{Claude Collin}
\author[a]{Julian Porras}
\author[a]{Nicolas Levraud}
\author[f]{Caroline Kulcs\'ar} 
\author[f]{Henri-Fran\c{c}ois Raynaud} 
\author[g]{Benoit Sassolas} 
\author[g]{Laurent Pinard} 
\author[g]{Christophe Michel} 
\author[a]{Cyrille Blanchard}
\author[a]{Ihsan Ibn Ta\"ieb} 
\author[b]{William Wang} 
\author[e]{Samir Maria}
\author[a]{Tristan Buey}
\author[b]{Vincent Chevalier}
\author[a]{Sylvestre Taburet}
\author[d]{Dalia Djetten}
\author[e]{Fran\c{c}ois Meyer} 
\author[a]{Pierre Baudoz} 
\author[a]{Elsa Huby} 
\author[h]{Sebastian Rabien}
\author[h]{Eckhard Sturm}
\author[h]{Richard Davies}

\affil[a]{LIRA, Observatoire de Paris, Universit\'e PSL, Sorbonne Universit\'e, Universit\'e Paris Cit\'e, CY Cergy Paris Universit\'e, CNRS, France}
\affil[b]{UNIDIA, Observatoire de Paris, Universit\'e PSL, CNRS, France}
\affil[c]{IPAG, CNRS, Universit\'e Grenoble Alpes, France}
\affil[d]{DT-INSU, CNRS, France}
\affil[e]{OSU THETA, CNRS, Universit\'e Bourgogne Franche Comt\'e, France}
\affil[f]{LCF, IOGS, CNRS, Universit\'e Paris Saclay, France}
\affil[g]{LMA, IP2I Lyon, CNRS, Universit\'e Claude Bernard Lyon 1, France}
\affil[h]{MPE, Germany}

\authorinfo{Further author information: send correspondence to Y. Cl\'enet, e-mail: yann.clenet "at" obspm.fr }

\begin{document} 
\maketitle

\begin{abstract}
MICADO is the designated ELT first light instrument, a near-infrared imager working at the diffraction limit of the telescope thanks to a SCAO and an MCAO mode. Since summer 2024, MICADO is in its AIT phase and all MICADO partners are currently integrating their subsystems. The MICADO SCAO module will follow three successive AIT phases in France: the $\beta$ flat configuration, the flat configuration and the final configuration. 

In this contribution I will first present the progress made in the AIT of the $\beta$ flat configuration and its closed loop results. I will then present the performed and on-going tasks for the flat and final configurations (manufacturing, delivery and integration of several SCAO subsystems and pieces of software) and the results obtained so far in the flat configuration. I will then conclude with the expected following steps of the SCAO module AIT in Garching.
\end{abstract}


\section{AN INTRODUCTION TO MICADO}
\label{sec:intro}  

The MICADO imager is the designated ELT first light instrument \cite{davies26}. Being available at telescope first light, MICADO will address all key topics of modern astronomy, from Solar system objects to the first galaxies of the Universe and including exoplanets. Working in the near-IR (0.8-2.4 $\mu$m) at the ELT diffraction limit, it will offer five observing modes:

\newpage
\begin{itemize}
\vspace{-3mm}
\setlength{\itemsep}{0pt}
\item Standard imaging: with 1.5 and 4 mas pixel scales, the corresponding FoV will be 19 and 51 arcseconds$^2$. 31 broad-band and narrow-band filters will be available.

\item Astrometric imaging: it drives MICADO design, with a gravity invariant implementation, a fixed mirror optical design, state-of-the-art ADC and dedicated
astrometric calibration and data pipeline.

\item High contrast imaging: it will use the central detector and will be enabled via a classical configuration of focal plane coronagraphs and Lyot stops\cite{baudoz26,huby26}, as well as pupil plane vAPP\cite{vondenBorn26} coronagraphs and sparse aperture masking. Pupil tracking will be available for angular differential imaging. 

\item Slit spectroscopy: it will provide coverage of a wide wavelength range simultaneously (J: 1.15-1.35 $\mu$m, HK: 1.52-2.45 $\mu$m or IzJ: 0.85-1.56 $\mu$m) at a resolution of about 20000 on faint compact or unresolved sources. Three slits will be available: 3\arcsec$\times$16 mas (IzJ), 15\arcsec$\times$20 mas (J \& HK), 3\arcsec$\times$48 mas (IzJ \& HK).

\item Integral field spectroscopy: it will offer 3D spectroscopy of a 0.45\arcsec$\times$0.39\arcsec field of view with a pixel size of 4 mas$\times$12 mas with a spectral resolution of 15000. It will cover H+K bands and possibly J-band.
\end{itemize}

\begin{figure}[t]
\begin{center}
   \begin{tabular}{c c}
 \includegraphics[height=6cm]{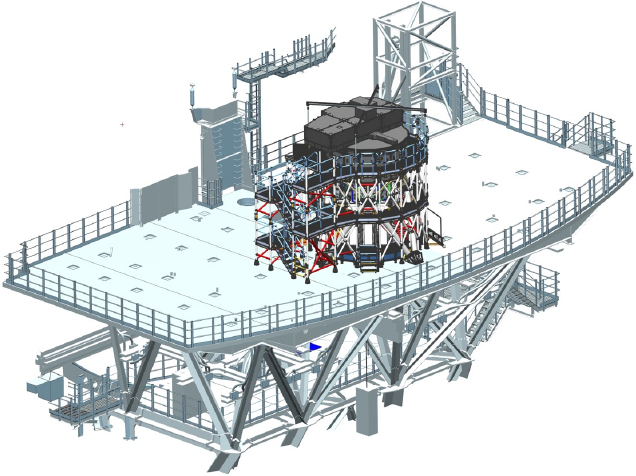} & \includegraphics[height=6cm]{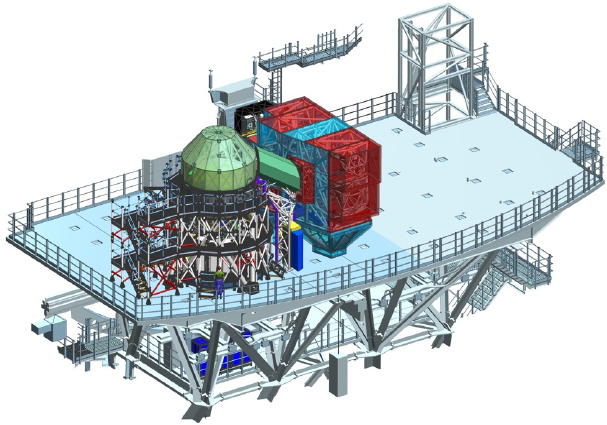}
       \end{tabular}
\end{center}
\caption{\label{fig:micado_standalone} Left: MICADO at the telescope in its standalone configuration, i.e. with the SCAO module alone and a dedicated optical relay. Right: MICADO at the telescope, coupled with MORFEO}
\end{figure}

In operations, MICADO will know two successive phases:
\begin{itemize}
\vspace{-3mm}
\setlength{\itemsep}{0pt}
\item the standalone phase (Fig.~\ref{fig:micado_standalone} left), starting at the ELT science first light, during which it will benefit  from the SCAO correction, developed within the consortium. A passive optical relay is developed by the consortium to feed the instrument. MICADO will be installed on the Nasmyth platform A of the ELT.
\item a so-called "M\&M" phase (Fig.~\ref{fig:micado_standalone} right), after the arrival of MORFEO at the telescope, few years after the ELT and MICADO first light. The instrument will then benefit from both the SCAO and a MCAO correction, the latter being developed by MORFEO together with an active optical relay \cite{ciliegi24}. In this phase, MICADO and MORFEO will be on the ELT Nasmyth platform B, requiring to move MICADO (including SCAO) from platform A to B. The SCAO module is not modified when MORFEO is installed: the space envelop, the interlock system, and the software configuration are  accounting for MORFEO  from the MICADO first light.
\end{itemize}

MICADO current planning is the following:
\begin{itemize}
\vspace{-3mm}
\setlength{\itemsep}{0pt}
\item 11/2018: PDR
\item 04/2021 -  07/2024: FDR reviews
\item 02/2029 - 10/2029: PAE process
\item 2030: commissioning
\item end of 2030: operations in standalone (w/ SCAO)
\item 2033: operations in M\&M configuration (w/ SCAO and MCAO)
\end{itemize}

\newpage
The project has officially validated its final design in July 2024 and is in its manufacturing, assembly, integration and tests phase. In practice, the FDR board and ESO agreed that these activities start from early 2023.  

\section{THE MICADO SCAO: MAIN SPECIFICATIONS AND FEATURES}
\begin{figure}[!t]
\begin{center}
 \includegraphics[height=13.4cm]{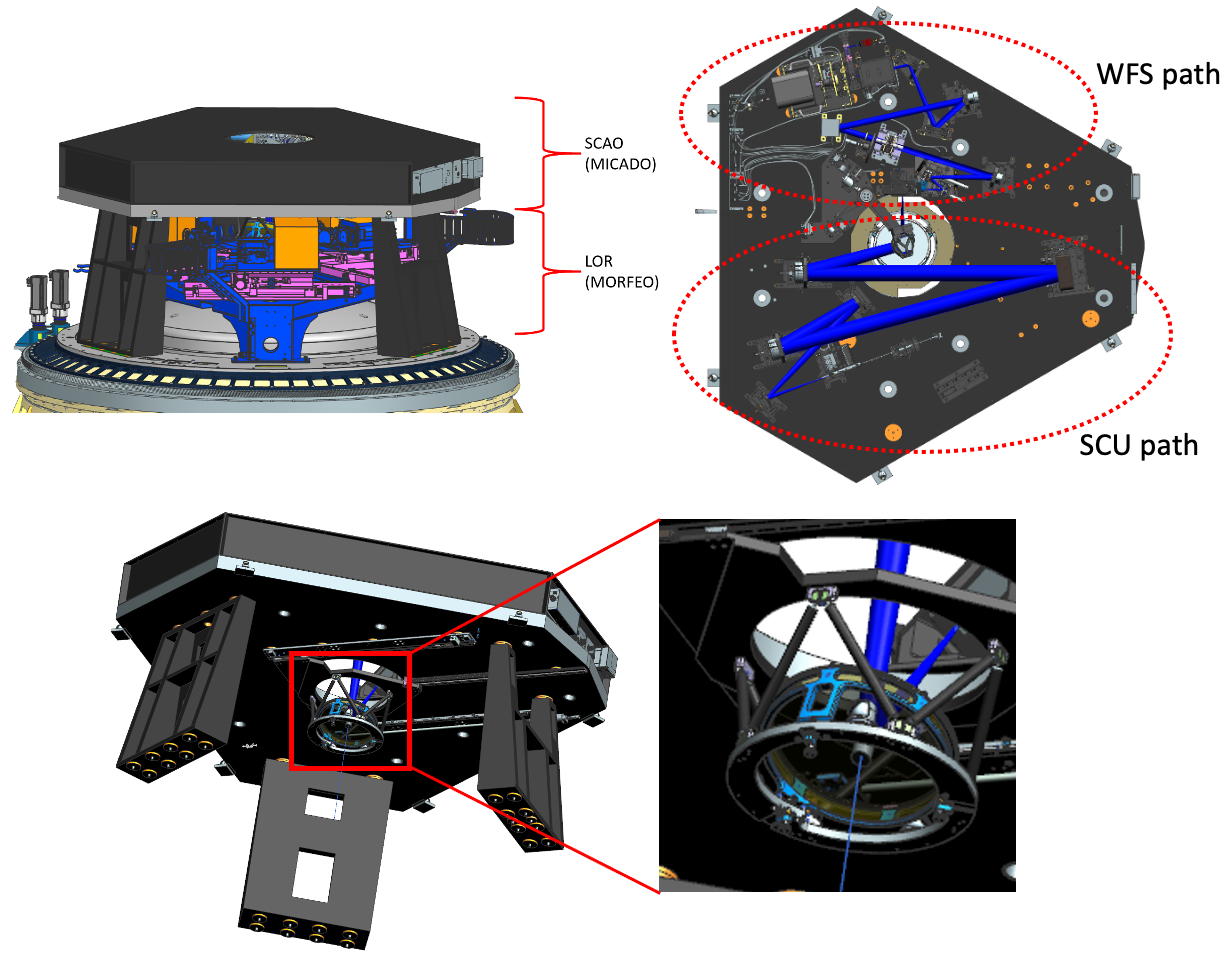} 
\end{center}
\caption{\label{fig:scao-description} Top left: side view of the subsystems inside the so-called Green Doughnut, mounted on top of the MICADO derotator+cryostat, namely the SCAO module at the top and the MORFEO LOR WFS at the bottom. Top right: top view of the SCAO bench in its on-sky configuration, showing the WFS path and the SCU path. Bottom: bottom views of the SCAO bench, showing the dichroic plate assembly.}
\end{figure}

The main sub-systems of the SCAO module are: its pyramid-based wavefront sensor (WFS), its calibration unit (so-called SCU), its dichroic plate assembly, its support structure, its real-time computer (hardware and software), its control software. Figure~\ref{fig:scao-description} shows different views of the SCAO module and these different sub-systems. 

The MICADO SCAO WFS is a pyramid-based WFS, sensing the light in the visible from 600 to 960 nm. The pyramid pupils will be imaged on the CCD220 detector of the ESO ALICE camera in 96$\times$96 pixels (leading to a subaperture size of 0.40 m). The total number of WFS measurements is then made of about 26000 elements. The pupil will be stabilized in position and clocking. The SCAO system will make use of the ELT M4 and M5 mirrors and the number of controlled modes will be from 2 to 4000. The loop will be running at up to 500 Hz. 

The SCAO system will be able to use a natural guide star with a R magnitude ranging from -1.5 (Betelgeuse) to 16. It will also be able to close the loop on extended objects, up to 1 arcsec in diameter (e.g. Solar system satellites) and will support differential tracking up to 100 arcsec/hour. The guide star will be selectable in a 6\arcsec$\times$20\arcsec\ patrol field.

For regular tuning of NCPA, health check at the beginning of the night and aging diagnostics, the SCAO module will include a dedicated calibration/maintenance unit with sources and a low order adapted ALPAO DMX37 mirror.

In terms of AO performance, the MICADO SCAO specification is 60\% of Strehl ratio at 2.2 $\mu$m, at 30$^\circ$ from zenith with a m$_R$=12 reference star, under medium seeing (0.702\arcsec), excluding windshake, vibrations and low wind effect.  

\section{UPDATES OF THE MICADO SCAO PERFORMANCE ESTIMATION}
\begin{figure}[t]
\begin{center}
   \begin{tabular}{c c}
 \includegraphics[height=4.4cm]{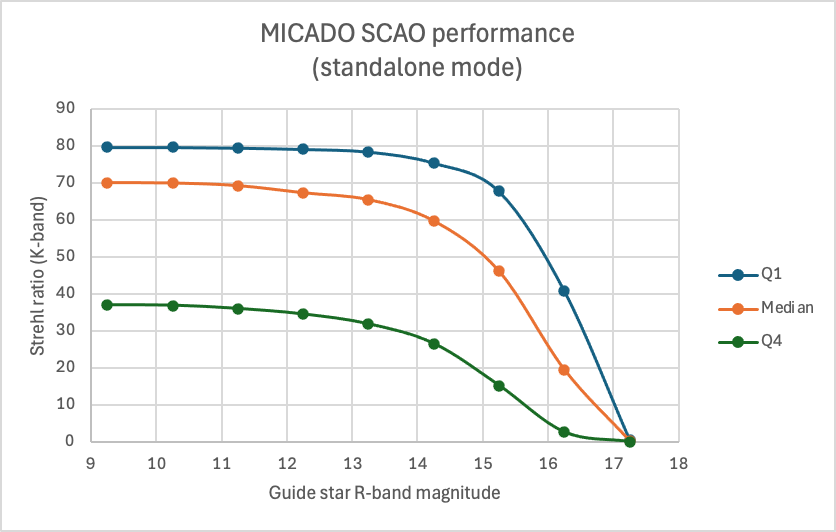} &  \includegraphics[height=4.4cm]{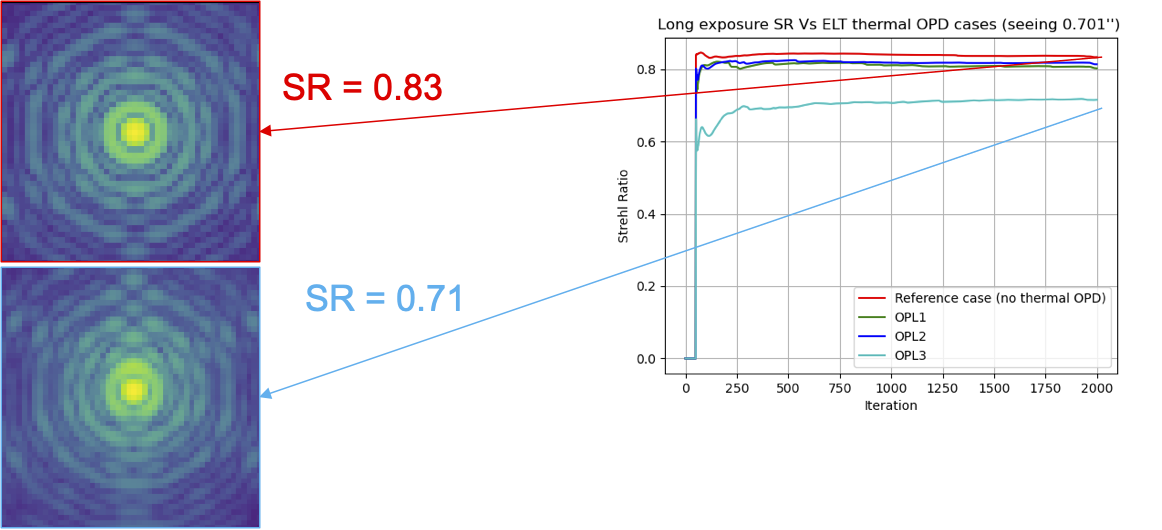}
       \end{tabular}
\end{center}
\caption{\label{fig:scaoperf} Left: Expected MICADO SCAO performance, accounting for its global error budget. Right: Evolution of the K-band Strehl ratio in COMPASS SCAO end-to-end simulations using either no thermal OPD, OPL1, OPL2 or OPL3. } 
\end{figure}

Figure~\ref{fig:scaoperf} at left shows the curves of the last update of the expected SCAO performance for the Q1, median and Q4 turbulence conditions, as defined ESO. They first account for the fitting error, the bandwidth error, the noise propagation error, the aliasing error and the atmospheric differential piston error, as estimated from AO end-to end simulations made with our \href{https://compass.pages.obspm.fr/website/}{COMPASS platform}. These AO simulations were done using an hybrid Linear Quadratic Gaussian (for tip-tilt) and classical integrator (for higher oder modes) AO control, considering the telescope windshake as simulated by ESO as well as three vibrations peaks, at 20 Hz, 60 Hz and 200 Hz, with a total amplitude of 42 nm rms. They also account for additional instrumental errors, such as manufacturing errors, mounting errors, alignement errors, etc.

In the past period, we have also assessed the impact of the dome seeing (or so-called "low wind effect") on SCAO performance. For that purpose we used the result of complex simulations performed by ESO\cite{holzo26}. These simulations are mixing computational fluid dynamics simulations with ray-tracing and 3D interpolation. The former were done in three M2 spiders structure configurations: partial cladding, full cladding and no cladding. As a result, we have been provided with 3 phase screens obtained for the full cladding case and representing the additional phase due to dome seeing at 3 successive time steps (270s, 330 s, 390 s). 

\begin{table}[!h]
   \centering
   \begin{tabular}{c|c|c|c|c|c} 
      Turbulence conditions    & Q1 & Q2 & median  & Q3 & Q4 \\
      \hline
      SCAO reference case (no dome seeing) & 0.91 & 0.86 & 0.83 & 0.78 & 0.56\\
      ELT thermal OPL1 & 0.90 & 0.84 & 0.80 & 0.76 & 0.53\\
      ELT thermal OPL2 & 0.90 & 0.85 & 0.81 & 0.77 & 0.54\\
      ELT thermal OPL3 & 0.81 & 0.75 & 0.71 & 0.67 & 0.49\\
   \end{tabular}
   \caption{\label{table:opl} K-band Strehl ratios obtained from SCAO COMPASS simulations, considering a 30$^\circ$ zenith angle and  various turbulence conditions, using the three ESO OPL phase maps.}
\end{table}

We ran COMPASS simulations using these three optical path length (OPL) maps as static additional phase maps, for the five turbulence conditions defined by ESO (at 30$^\circ$ zenith angle): Q1, Q2, median, Q3 and Q4. The resulting Strehl ratios obtained in the K-band are given in Table~\ref{table:opl}. It shows that performance can be penalized by up to 10\% due to petalling in worst case conditions (OPL3), even in the best telescope configuration (full cladding). Illustration of the long exposure Strehl ratio evolution as well as final PSF are given in Fig.~\ref{fig:scaoperf} right.

\section{THE $\beta$ FLAT CONFIGURATION}
\begin{figure}[t]
\begin{center}
   \begin{tabular}{c c }
 \includegraphics[height=5cm]{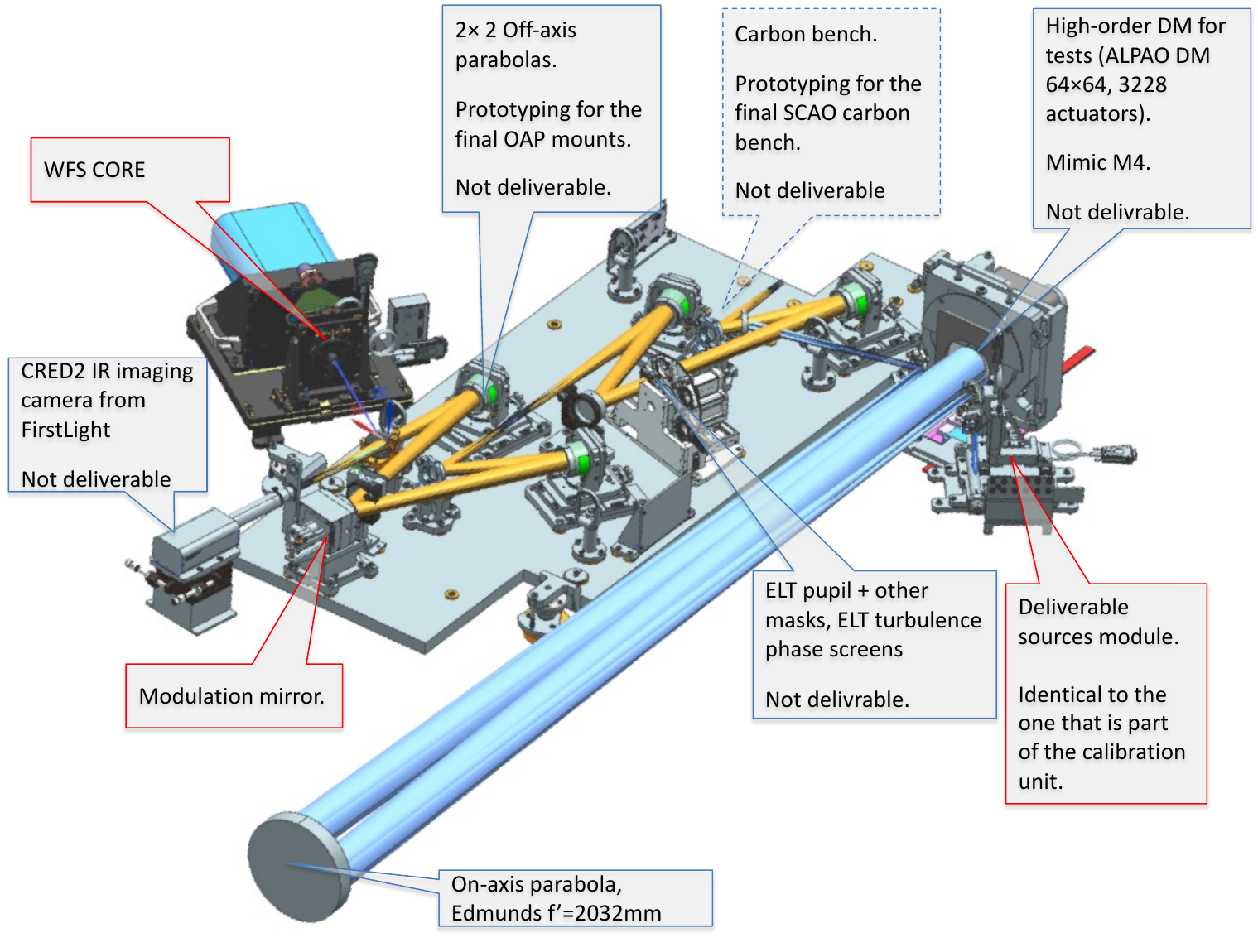} & \includegraphics[height=4.5cm]{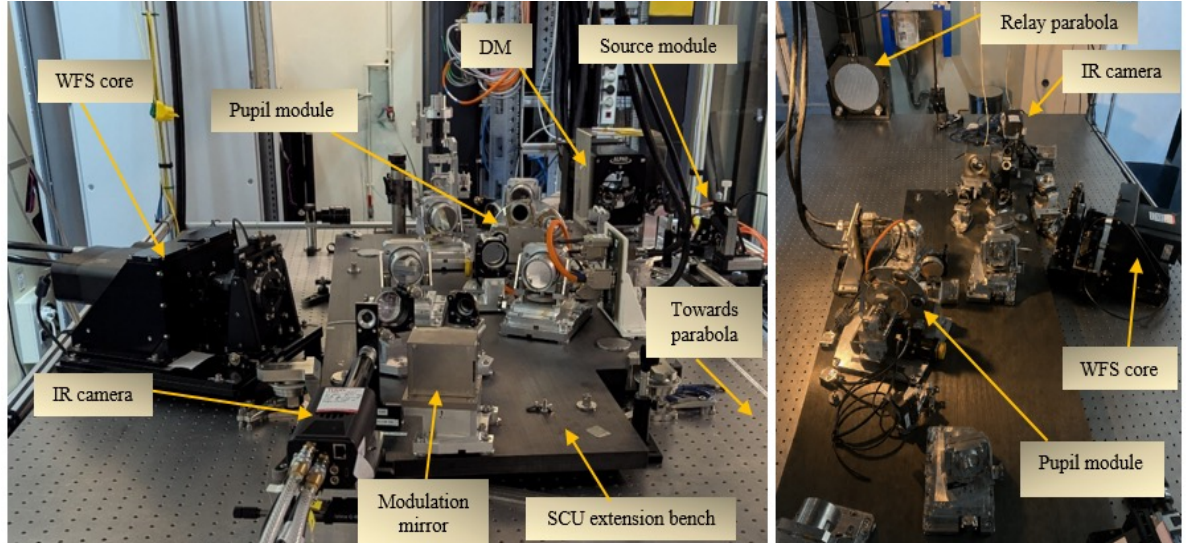} 
       \end{tabular}
\end{center}
\caption{\label{fig:betaflat} Left: CAD view of the $\beta$ flat configuration, commenting the different subsystems. Middle and right: pictures of the $\beta$ flat configuration.}
\end{figure}

The $\beta$ flat configuration is the first of the three phases of the MICADO SCAO AIT in France. It took place in the "ATLAS" clean room at LIRA at Observatoire de Meudon (Fig.~\ref{fig:betaflat}). It started mid 2023 and stopped in April 2026.

The first goal of this configuration was to develop and validate our telescope and turbulence simulator. It is made of a source module, a pupil module, a high order deformable mirror (HODM), all mounted on a carbon bench. 

\begin{figure}[b]
\begin{center}
   \begin{tabular}{c c }
 \includegraphics[height=3.6cm]{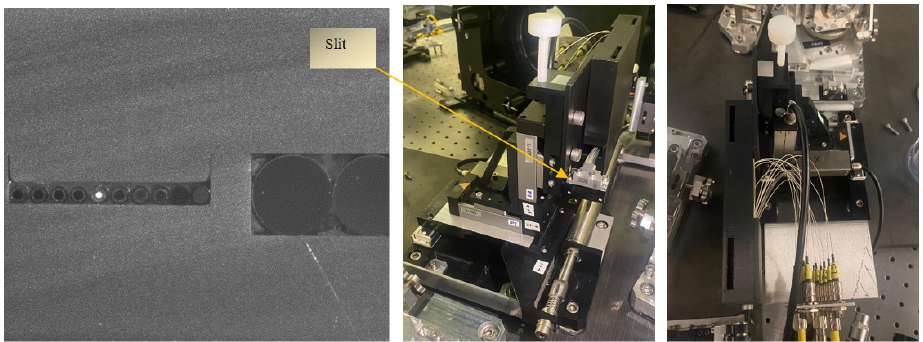} & \includegraphics[height=3.6cm]{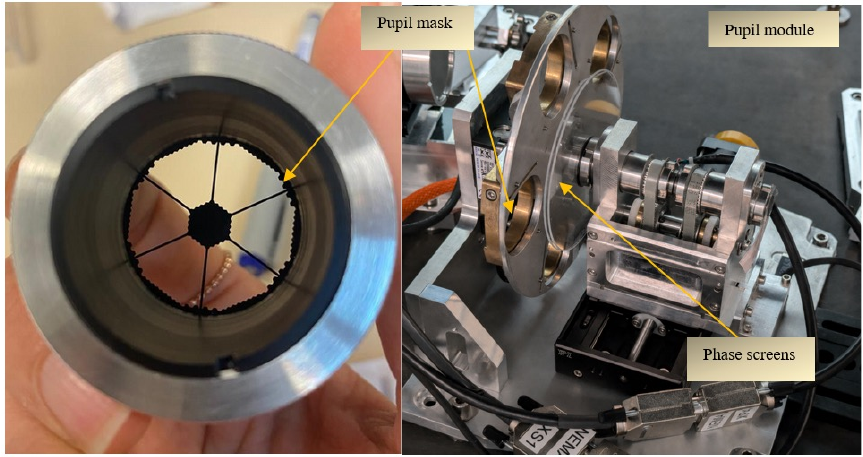} 
       \end{tabular}
\end{center}
\caption{\label{fig:module} Leftern three pictures: source module pictures showing the arrangement of the fibers and  two side views of the assembly. Rightern two pictures: pictures of the ELT pupil mask and of the pupil module assembly}
\end{figure}

The source module consists in twelve optical fibers arranged along a slit (Fig.~\ref{fig:module}). These fibers are single-mode fibers for diffraction-limited source generation, multimode fibers for seeing-limited simulations, large-core fibers (up to 1 mm diameter), visible fibers dedicated to wavefront sensor testing, and infrared fibers for the imaging channel characterization. The source module is motorized so that it is possible to remotely choose the fiber to be injected in the optical path, as well as to adjust laterally (with PI L505 linear stages) and in focus (with a PI M122 translation stage) the fiber position. The fibers are coupled to LEDs, laser sources and other emitters like a plasma lamp that are integrated into a dedicated enclosure, so-called "lamp box". The complete assembly is remotely controlled with a custom-made electronics board that allows to select the source and its intensity.

\begin{figure}[t]
\begin{center}
   \begin{tabular}{c c c c }
 \includegraphics[height=3cm]{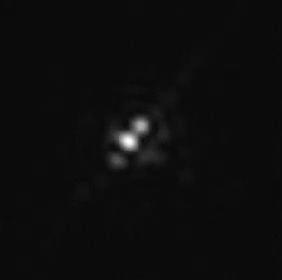} & \includegraphics[height=3.2cm]{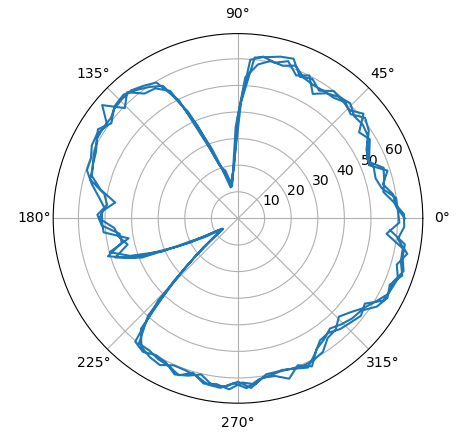} & \includegraphics[height=3.2cm]{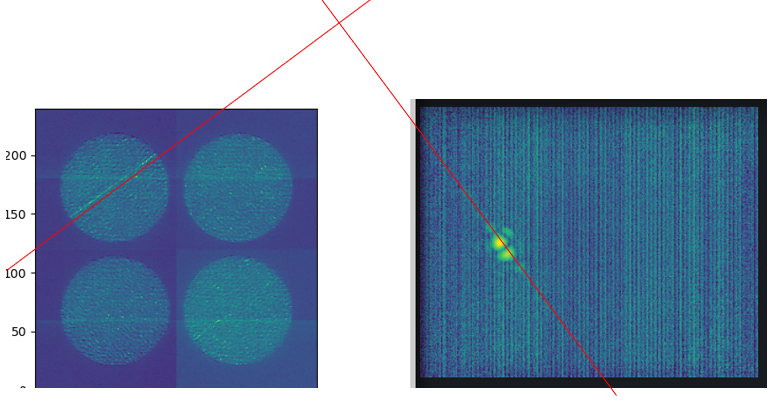} & \includegraphics[height=3cm]{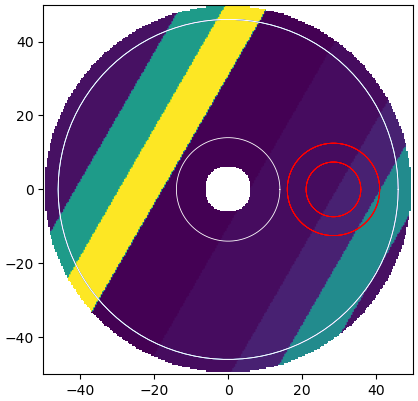}
       \end{tabular}
\end{center}
\caption{\label{fig:petalling} From left to right: snapshot of our IR camera showing a petalling effect, measured Strehl ratio  at 1550 nm wrt the rotation angle of the most affected phase screen, WFS camera and IR camera images showing the correspondance between the elongation direction of IR PSF and the direction of a line observed on a WFS pupil image, mean levels of phase on the most affected phase screen (green is around 300 nm, yellow is around 700 nm, blue is around 0 nm, the red circles are the beam footprint in the $\beta$ flat configuration for the smaller and in the flat configuration for the larger).}
\end{figure}

The pupil module supports two kinds of equipment (Fig.~\ref{fig:module}). First it holds masks optically conjugated with the telescope pupil. Hence it holds a mask with the ELT pupil shape, though with the older 51 cm spiders. It also holds a mask with a 38.542 m (ELT pupil mean diameter) circular unobstructed shape and additional masks dedicated to alignment purposes. The pupil module also holds two turbulence phase screens to simulate the atmospheric turbulence. First close loop tests demonstrated that the resolution of the original phase screens were too coarse, leading to too large discontinuity steps and then a bad turbulence simulation. This resolution, with pixel of 100 $\mu$m, was though about the best achievable for an acceptable price when we bought them in 2020 ($\approx$40 k\euro). We then bought new phase screens from a second manufacturer, fortunately able to provide us with phase screens at a much better resolution (typically a few microns), still at a decent price. However, their tests after delivery demonstrated a petalling like behavior. After some investigation, we understood it was coming from a normalization by bands done by the manufacturer on the data we had sent to him, resulting in an array with bands and large phase steps between them (Fig.~\ref{fig:petalling}). We then had  to remanufacture them, taking care of a proper normalisation by the manufacturer.

The second goal of the $\beta$ flat configuration was to close the AO loop with the minimal set of final SCAO subsystems. Hence, as already reported at the last SPIE Astronomical telescope + instrumentation conference\cite{clenet24}, we  procured early in advance the pyramid optics and mount,  the modulation system as well as the ALPAO HODM (meant to simulate the ELT M4). We had already developed the HRTC SW\cite{sevin24} and deployed the very first releases of the RTC Toolkit developed by ESO. We then developed AO-oriented pieces of software:
\begin{itemize}
\item Pupil-fitting algorithms, tested against turbulence, flux level, pupil position and rotation, so that we reached a 0.1\% pupil diameter precision 
\item Photometric calibration of the sources
\item HODM characterization (unknown influence functions)
\item Fast computation of synthetic control matrix
\item Modal basis computation accounting for actuators outside the pupil and specific AIT config
\item Estimation of DM mis-registrations and optical gains
\item Phase diversity
\end{itemize}

We also developed AIT custom pieces of software, with GUI interfaces, to be able to control the bench and to monitor the status of the HRTC and SRTC processes (Grafana, Loki, RTC webservices)\cite{sevin26}. We also initiated preliminary on-bench tests of the LQG+integrator AO control. And building up from all these developments, we have been able to close the AO loop, reaching 88\% Strehl ratio at 1550 nm. 

\section{THE FLAT CONFIGURATION}
The flat configuration is the second of the three phases of the MICADO SCAO AIT in France. It takes place in the "ATLAS" clean room at LIRA at Observatoire de Meudon. It started in August 2025 and is planned to stop in Autumn 2026.

The fist goal of the flat configuration is to develop, integrate and align almost all WFS and SCU subsystems: WFS pupil viewer (to get a pupil image at very high resolution), WFS pupil imaging lens (to adjust the pupil laterally), WFS K-mirror (to adjust the pupil in clocking), NCPA wheel (to support phase screens correcting static NCPA), WFS field selector (to pick up and track the possibly moving AO reference source), WFS ADC, WFS and SCU off-axis parabola/flat mirrors mounts. Apart from the ADC optics and mechanics, we have received all the optics of these subsystems, manufactured all the corresponding support mechanics and mounted the optics in their mechanical supports. This mounting has been made with the objective to get the smallest wavefront error (WFE) for each optics. Hence, it followed a systematic iterative mount - interferometric measure - adjust process to avoid any mechanical stress on the optics. We adjusted some mounting for example by using additional washer to reduce spring preload. We also used blade bonding technique with blades in Invar in order to minimize thermo-elastic effects and to provide a near-isostatic mounting scheme. It allowed us to significantly reduce the overall WFE. In the end, for the WFS optics, the total WFE measured using a ZYGO interferometer is 36 nm rms. This WFE does not account for the field lens, the NCPA phase plate nor the ADC. With a total specified budget of 60 nm rms, we then have room for these missing WFS optics.

\begin{figure}[t]
\begin{center}
   \begin{tabular}{c c }
 \includegraphics[height=3.8cm]{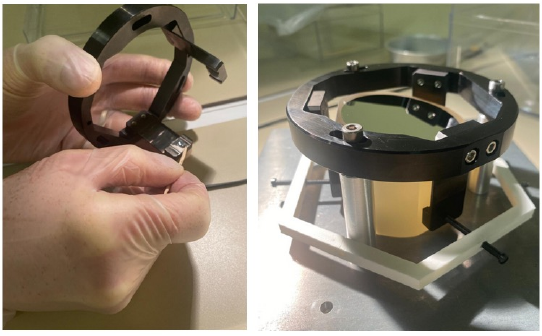} & \includegraphics[height=3.7cm]{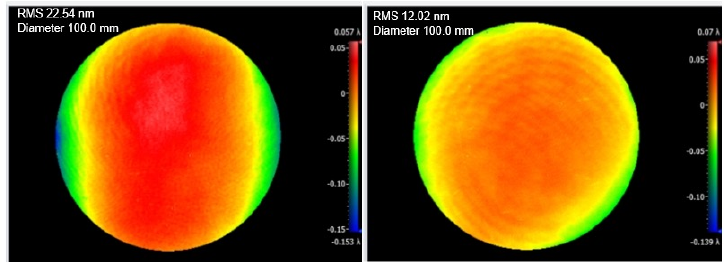} 
       \end{tabular}
\end{center}
\caption{\label{fig:flatalign} Left: illustration of bonding blade mounting of one of the off-axis parabola. Right: illustration of the reduction of the WFS fold mirror \#1\ WFE from 23 nms to 12 nm rms by reducing the spring preload after the addition of a washer.}
\end{figure}
\begin{figure}[b]
\begin{center}
   \begin{tabular}{c c }
 \includegraphics[height=5.2cm]{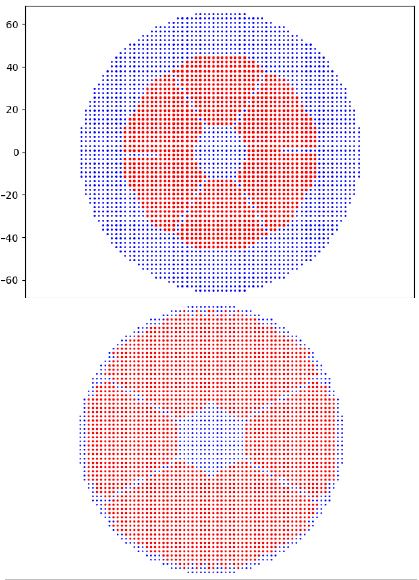} & \includegraphics[height=5.2cm]{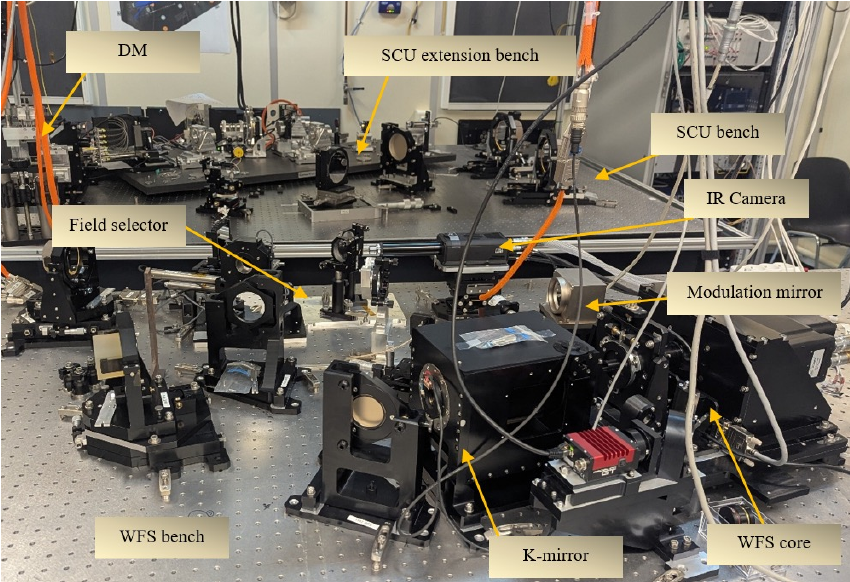} 
       \end{tabular}
\end{center}
\caption{\label{fig:flatconf} Left: comparison of the useful HODM actuators (in red) between the $\beta$ flat configuration and the flat configuration. Right: Picture of the flat configuration.}
\end{figure}

Then these WFS and SCU optics have been mounted and aligned. We followed for that purpose our carbon bench strategy. Indeed, our final SCAO bench is made of carbon and not aluminium, preventing from positioning the optics supports and the alignement tools wherever we want. Therefore, we made a first alignment on aluminium benches with the mounts at mechanical positions and with the alignment tools where needed. We confirmed that the alignement was possible using only the limited adjustment capabilities of each mount. And we plan to transfer to the final bench the set-up as it is, without any new adjustments. We validated this strategy during the $\beta$ flat configuration with the SCU extension bench, also made in carbon, and with the optics mounted on this bench. More details on these optics mounting and alignement procedures  are given in Chemla et al. (2026)\cite{chemla26}. 

The second goal of the flat configuration is to develop and deploy the  instrument control software corresponding to these WFS and SCU optomechanical subassemblies (from low-level to high level), first with custom-made AIT software, then with the ESO control software infrastructure releases. It represents twelve devices to control (including five custom devices), with ESO function control systems (FCS), one technical camera to read through the ESO camera control framework, a control software interface with the SCAO RTC through a custom API and a control software interface with the MICADO instrument through a dedicated API. All the control software is currently using our custom-made AIT software and we are progressively migrating each subsystem to the ESO control software. 

\begin{figure}[t]
\begin{center}
   \begin{tabular}{c c }
 \includegraphics[height=5.5cm]{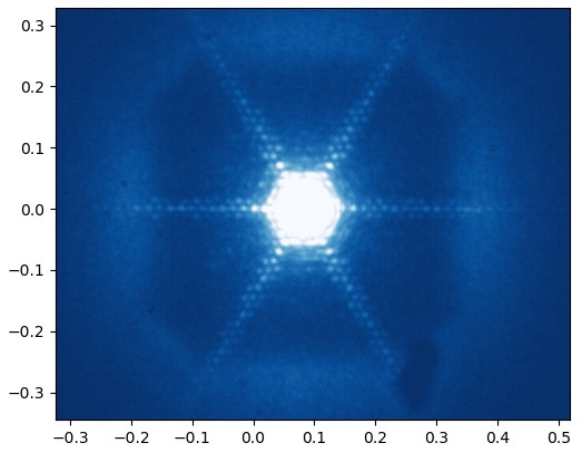} & \includegraphics[height=4.5cm]{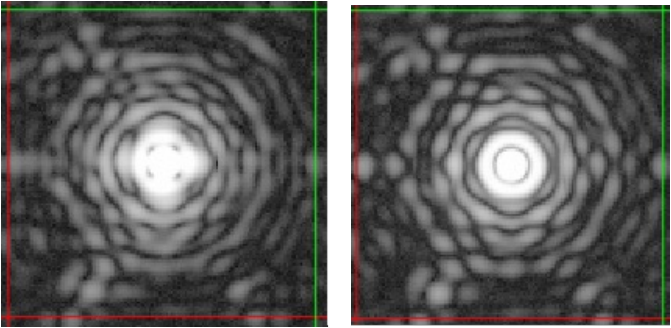} 
       \end{tabular}
\end{center}
\caption{\label{fig:flatconf_closed} Left: Image of the corrected PSF in flat configuration. Right: PSF without and with phase diversity NCPA correction.}
\end{figure}

The final goal of the flat configuration is to close the AO loop with these final subsystems using the ESO control software. When switching from the $\beta$ flat configuration to the flat configuration in April 2026, 	and despite the significant changes between the two configurations (Table~\ref{table:compconfig}, Fig.~\ref{fig:flatconf} left), we have been able to rapidly close the loop, using our custom AIT software, thanks to all the software developments made in the $\beta$ flat configuration. Then, we  obtained about 33\% Strehl ratio at 1550 nm in the presence of turbulence (which is about what is expected from simulation) and started to optimize the already developed AO-oriented pieces of software like phase diversity techniques (Fig.~\ref{fig:flatconf_closed}).

\begin{table}[!h]
   \centering
   \begin{tabular}{c|c|c} 
      Parameters    & $\beta$ flat configuration & flat configuration \\
      \hline
      pupil diameter on phase screens & 14.8 mm & 25 mm\\
      pupil diameter on HODM & 66 mm & 93 mm\\
      number of actuators across pupil diameter & 44 & 62\\
      number of controlled modes & 1200 & 2500\\
      Value of $D/r_0$ with 2 phase screens & 226 & 382\\
      ELT-equivalent seeing at 0.5 $\mu$m & &\\
      high altitude screen & 0.37\arcsec & 0.63\arcsec (Q2)\\
      low altitude screen & 0.43\arcsec & 0.72\arcsec (median)\\
      both & 0.60\arcsec & 1.02\arcsec (Q3/Q4)\\
        \end{tabular}
   \caption{\label{table:compconfig} Comparison of the main configuration parameters between the $\beta$ flat configuration and the flat configuration.}
\end{table}

\pagebreak

\begin{figure}[t]
\begin{center}
   \begin{tabular}{c c }
 \includegraphics[height=4cm]{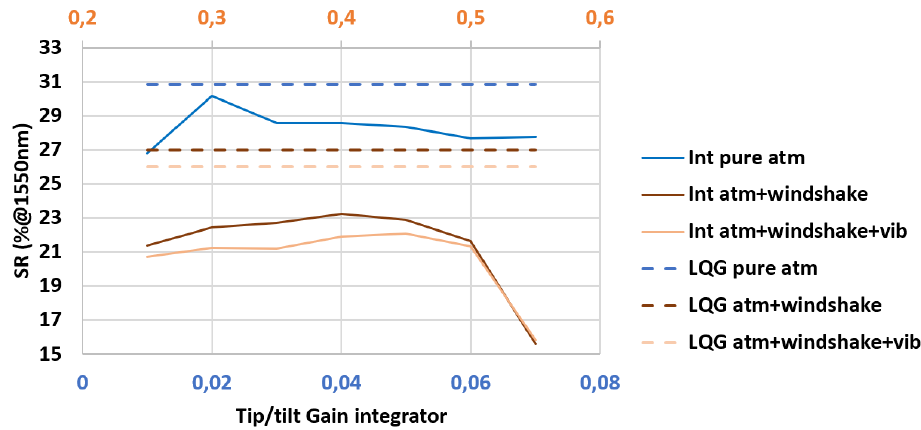} & \includegraphics[height=4cm]{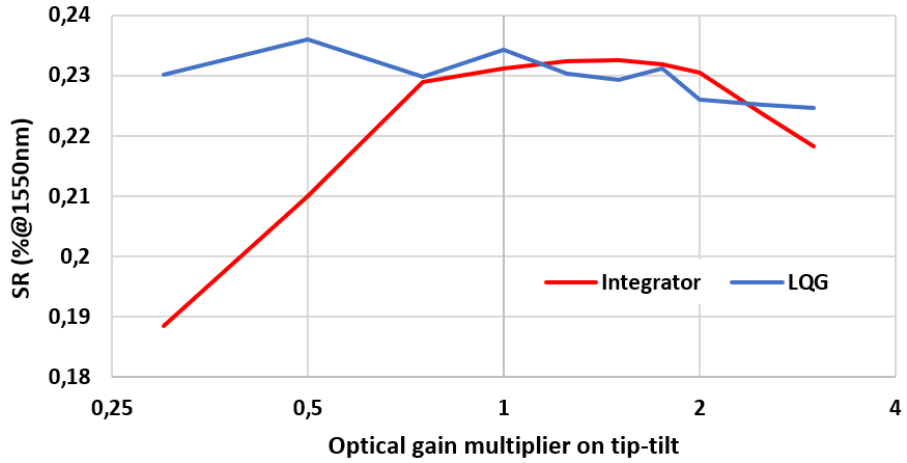} 
       \end{tabular}
\end{center}
\caption{\label{fig:lqg} Left: AO performance for pure integrator (plain lines) and LQG+integrator (dashed lines) AO control in the case of pure atmosphere (blue), atmosphere and windshake (red) and atmosphere, windshake and vibrations (orange) as measured on bench in the flat configuration. Right: on-bench evaluation of the pure integrator and hydrid LQG+integrator robustness wrt to misestimation of the pyramid optical gain.}
\end{figure}

We also continued the LQG+integrator on-bench tests and demonstrated the gain in performance brought by this hybrid AO control in the presence of windshake and vibrations\cite{levraud26}. For these tests, windshake was introduced as perturbations on our slow tip-tilt mirror, that mimicks M5, while vibrations were introduced as perturbations on our HODM. For the tip-tilt correction, a temporal separation was performed between the slow tip-tilt mirror and the HODM. Comparison of performance between the pure integrator and the hybrid LQG+integrator is shown in Fig.~\ref{fig:lqg} left. We also confirmed the simulation results about the robustness of the hybrid AO control with respect to a misestimation of the pyramid optical gain (Fig.~\ref{fig:lqg} right).

\section{THE FINAL CONFIGURATION}
The final configuration is the last of the three phases of the MICADO SCAO AIT in France. It takes place in the "Les Communs" integration hall at Observatoire de Meudon. It started in March 2026 and is planned to stop in August 2028, with the delivery of the SCAO module to MPE for the MICADO system integration.

The first goal of this final configuration is to develop and validate the remaining WFS and SCU subsystems: the SCAO carbon bench, the SCAO dichroic, the SCU deployment arm and the SCU deformable mirror.

\begin{figure}[b]
\begin{center}
 \includegraphics[height=4.1cm]{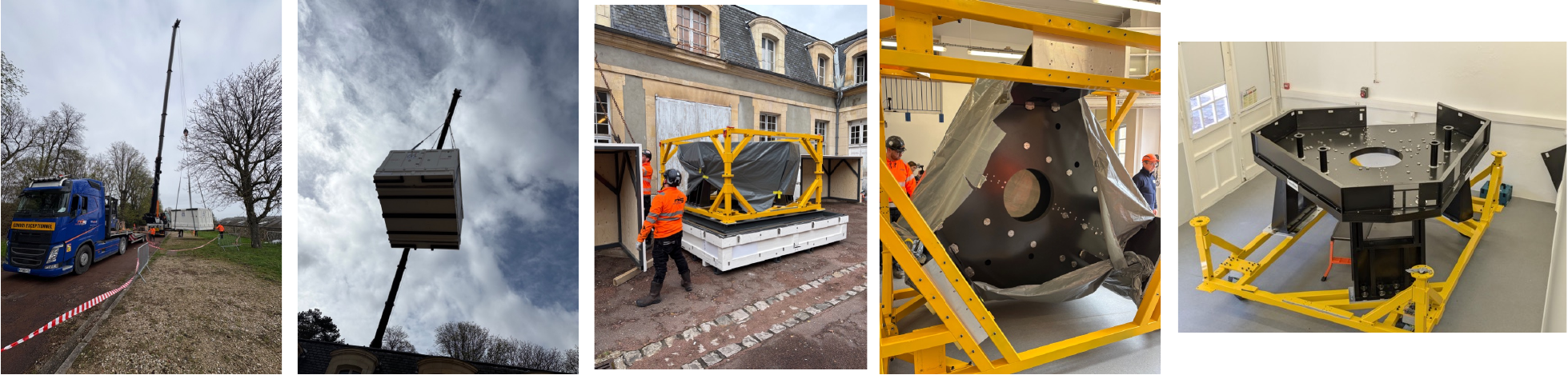} 
\end{center}
\caption{\label{fig:bench} Pictures of the delivery of the final SCAO carbon bench in the "Les Communs" integration hall at Observatoire de Meudon on March 16$^{th}$.}
\end{figure}

Concerning the SCAO carbon bench, manufactured by the company CarbonVision in Germany, it has been delivered to Meudon on March 16$^{th}$ this year, with the help of the CNRS ULISSE service. First, the maintenance trolley of the bench has been transported from its manufacturer MGTL to the manufacturer of the box E3Cortex (all in the suburbains of Paris) to be mounted in it. Then the assembly has been transported to the manufacturer of the bench in Germany. There the bench has been  mounted in its trolley and all together installed in the box. Everything then went back to the Observatoire de Meudon. Then, we had  to crane over the historical building these about 3 tons of hardware made of the bench (430 kg), its trolley (750 kg) and their transportation box (1800 kg). When arrived in front of the integration hall, we opened the box, lifted out the trolley and the bench. Since the integration hall doors are too narrow, we had to flip by 90$^\circ$ the trolley and the bench before entering into the integration hall and to flip it back horizontally once inside. We then have controlled the stringent specification of 0.1 mm for the absolute location of the inserts used to position the SCAO subsystems on the bench. Using a FARO arm, we measured  the inserts positions and confirmed that they are all within 0.06 mm with respect to their defined positions. The bench planarity (upper and lower surface) has been controlled by CarbonVision to be better than 0.05 mm/m. The accuracy of the distance between the upper surface of the bench and the MICADO interface has been measured to be better than 0.01 mm.
 
 \begin{figure}[t]
\begin{center}
 \includegraphics[height=4cm]{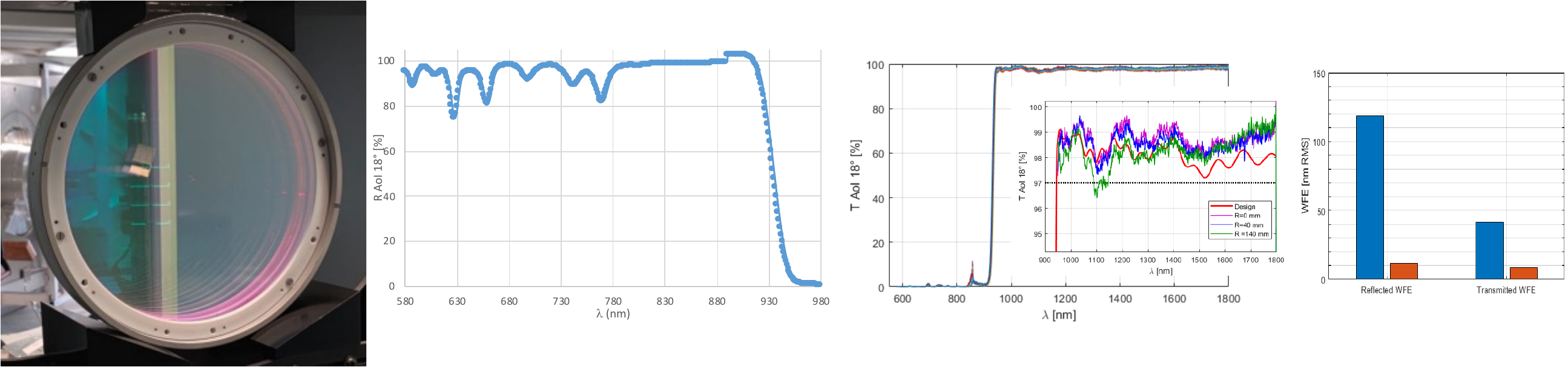} 
\end{center}
\caption{\label{fig:dichroic} From left to right: picture of the coated dichroic plate at LMA, measured transmittance at 18$^\circ$ inclination, measured reflectance at 18$^\circ$ inclination (with an inset showing the transmittance as designed and as measured at three positions), transmitted and reflected measured and specified WFE.}
\end{figure}

Concerning the SCAO dichroic, the 305 mm diameter CaF$_2$ plate has been procured from the Hellma company and then polished by Bertin Winlight. It has then been coated by the French laboratory LMA/IP2I\cite{sassolas26}. The coating consists in 70 layers of Ti;Ta$_2$O$_5$ and SiO$_2$, for a total thickness of 7.5 $\mu$m. An anti-reflective coating, made of 16 layers and 2 $\mu$m thick, was also deposited on the backside to mitigate the Fresnel reflection loss in the near-infrared. The average transmittance between 960 and 1800 nm, at 18$^\circ$ inclination,  is 98.3\%, while 97\% were specified. The average reflectance between 600 and 960 nm, at 18$^\circ$ inclination, is 89.3\%, while 92\% were specified. This gap, coming from dips in the transmittance curve whose origin is still under investigation, is though completely acceptable. In terms of WFE, measurements after coating show:
\begin{itemize} 
\item in reflection, 118.8 nm rms over 150 mm diameter and 11.8 nm rms on average over 13 subpupils of 40 mm in diameter, while specifications were 200 nm rms and 20 nms rms respectively
\item in transmission, 41.3 nm rms over 290 mm diameter and 8.2 nm rms on average over 42 subpupils with 40~mm in diameter, while specifications were 150 nms rms and 15 nm rms respectively.
\end{itemize}

\begin{figure}[b]
\begin{center}
 \includegraphics[height=4cm]{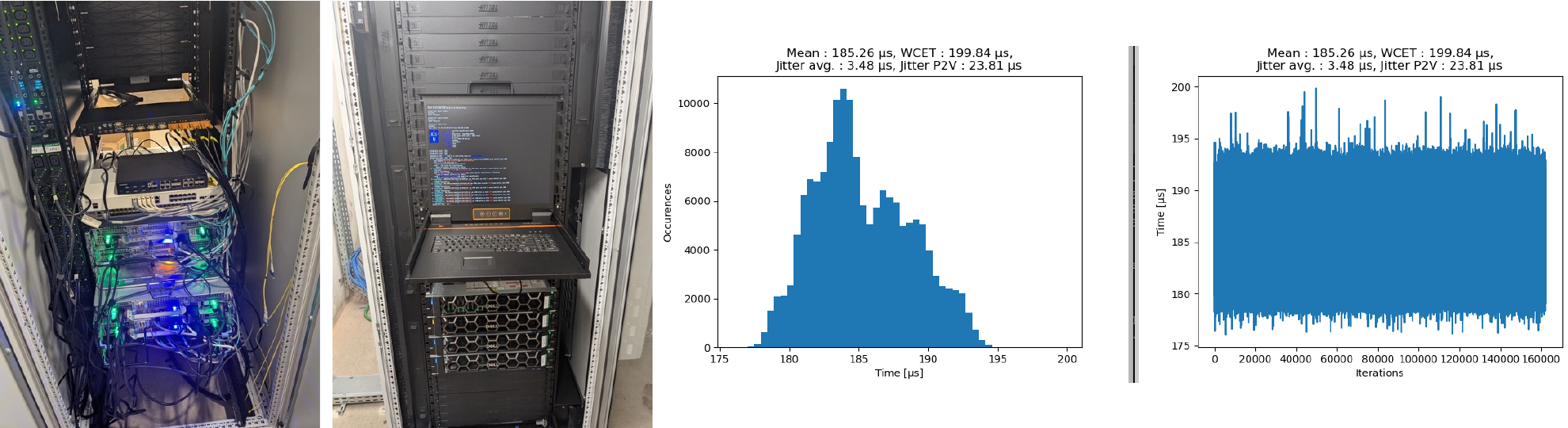} 
\end{center}
\caption{\label{fig:rtc} From left to right: pictures of final RTC servers in their AIT cabinet. Performance measurements in terms of latency and jitter.}
\end{figure}

\pagebreak

Regarding the SCU deformable mirror, the final delivery of this modified ALPAO DMX37 mirror has not been done: despite the fulfill of the best flat specification, ALPAO is currently investigating an issue of stroke for few actuators.

The second goal of this configuration is to finalize the development and deployment of the complete instrument control software (both low-level and high level), of the electronics cabinets and of the complete cabling. These tasks are currently on-going.

A third goal of this configuration is to deploy the complete RTC hardware and software\cite{ferreira26}. Concerning the hardware, it has been already procured: following the ESO IT standards, the HRTC and SRTC compute nodes are Dell PowerEdge R760XA servers with 2 GPU NVIDIA H100 NVL each and the HRTC and SRTC gateways are Dell PowerEdge R7625 servers. Performance tests in terms of latency, jitter and robustness (number of iterations without any crash) have proven to be within specifications. Development of the final software, particularly following the successive releases of the ESO RTC toolkit is on-going.

The final goal of this configuration is to validate the full SCAO subsystem before SCAO delivery to MPE in August 2028 for MICADO system integration.  For that purpose we will perform the SCAO acceptance with all the opto-mechanical subsystems, the final control software, the final RTC hardware and software, the final cabling and electronics cabinets. Though, we might decide to mount the final SCAO dichroic only at MPE (to avoid too much handling of this fragile optics). And then, there will miss only one piece of optics: the final phase plates for NCPA. The measurement of the MICADO NCPA will be done only at MPE of course, so that the manufacturing of this phase plate will occur only then.

\begin{figure}[!t]
\begin{center}
 \includegraphics[height=5cm]{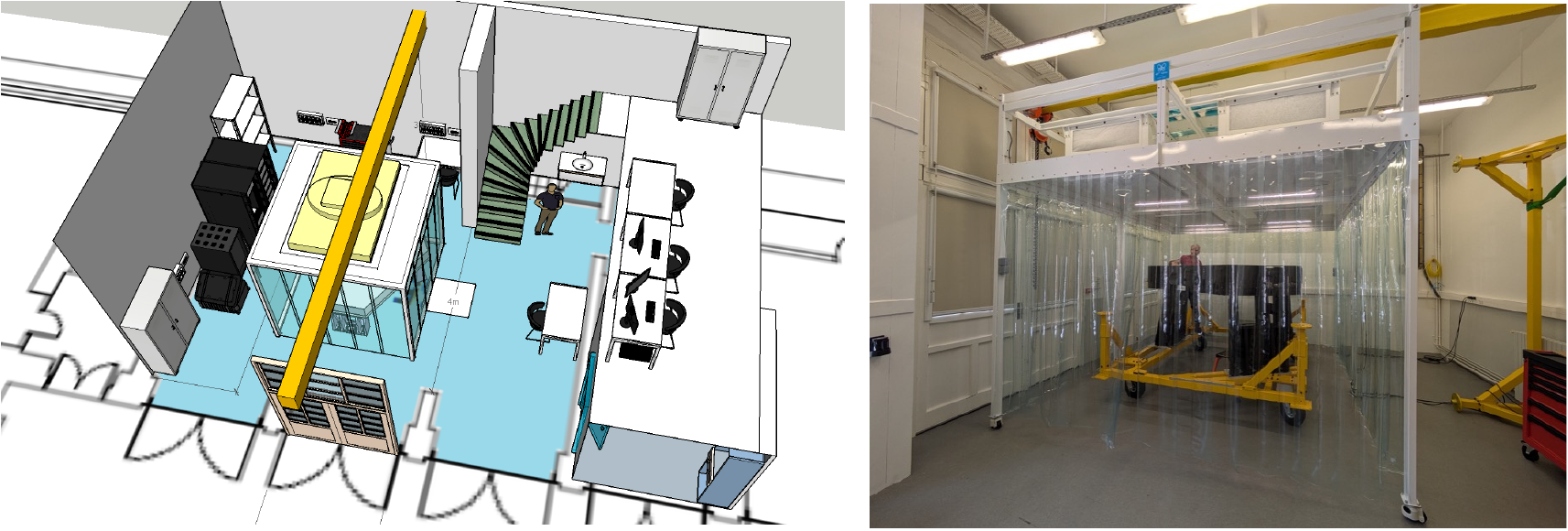} 
\end{center}
\caption{\label{fig:communs-description} Left: CAD view of the "Les Communs" integration hall. Right: Picture of the same hall, with the recently delivered cleaning tent, with the SCAO carbon bench and its trolley underneath.}
\end{figure}

\section{PERSPECTIVES}
For the SCAO module, the system integration will start at the MPE X4 integration hall. It will be mounted on the MICADO derotator and cryostat during the cold phase of the MICADO MAIT. After a bit less than a year, the SCAO module will be moved to the ESO Large Integration Hall in Garching for the remaining integration tests and the Preliminary Acceptance in Europe.

\acknowledgments 
 
This work has benefited from the support of 1) the French Programme d'Investissement d'Avenir through the project F-CELT ANR-21-ESRE-0008, 2) 2) the CNRS 80 PRIME program, 3) the CNRS INSU IR budget, 4) the Action Sp\'ecifique Haute R\'esolution Angulaire (ASHRA) of CNRS/INSU co-funded by CNES, 5), the Observatoire de Paris, 6) the Ile de France region (DIM ACAV/ACAV+ and ORIGINES) and 7) the LIRA and LCF laboratories

\bibliography{report} 
\bibliographystyle{spiebib} 

\end{document}